\documentclass[aps,prx,twocolumn,superscriptaddress]{revtex4-2}

\usepackage{graphicx}
\usepackage{wrapfig}
\usepackage{setspace}
\usepackage{etoolbox}
\usepackage{amsmath}
\usepackage{xcolor,soul}
\usepackage{float}

\makeatletter
\patchcmd{\frontmatter@abstract@produce}
{\vskip200\p@\@plus1fil
	\penalty-200\relax
	\vskip-200\p@\@plus-1fil}
{}
{}
{}
\makeatother

\begin{document}

\title{Landau theory applied to antiferroelectric ordering in ferroelectric nematic liquid crystals}
	
\author{Manisha Badu}
\thanks{These authors contributed equally to this work.}
\affiliation{Department of Physics, Kent State University, Kent OH 44242}

\author{Arjun Ghimire}
\thanks{These authors contributed equally to this work.}
\affiliation{Department of Physics, Kent State University, Kent OH 44242}

\author{Milon}
\thanks{These authors contributed equally to this work.}
\affiliation{Materials Science Graduate Program and Advanced Materials and Liquid Crystal Institute, Kent State University, Kent, OH 44242}

\author{Priyanka Kumari}
\affiliation{Materials Science Graduate Program and Advanced Materials and Liquid Crystal Institute, Kent State University, Kent, OH 44242}

\author{Hari Krishna Bisoyi}
\affiliation{Advanced Materials and Liquid Crystal Institute, Kent State University, Kent, OH 44242}

\author{Oleg D. Lavrentovich}
\affiliation{Department of Physics, Kent State University, Kent OH 44242}
\affiliation{Materials Science Graduate Program and Advanced Materials and Liquid Crystal Institute, Kent State University, Kent, OH 44242}

\author{James Gleeson}
\affiliation{Department of Physics, Kent State University, Kent OH 44242}

\author{Antal Jakli}
\affiliation{Department of Physics, Kent State University, Kent OH 44242}
\affiliation{Materials Science Graduate Program and Advanced Materials and Liquid Crystal Institute, Kent State University, Kent, OH 44242}

\author{Samuel Sprunt}
\email[]{ssprunt@kent.edu}
\affiliation{Department of Physics, Kent State University, Kent OH 44242}
\affiliation{Materials Science Graduate Program and Advanced Materials and Liquid Crystal Institute, Kent State University, Kent, OH 44242}

\date{\today}
	
\begin{abstract}
	The polarization and mass density modulation associated with antiferroelectric ordering in two ferroelectric nematic liquid crystals, a prototypical single compound (referred to as DIO) and a commercial mixture (FNLC919), are studied experimentally as a function of temperature. The wavenumber of the polarization modulation, $q_A$, is determined by small angle X-ray diffraction from the weak, accompanying density wave (wavenumber $q_S = 2 q_A$) using a conventional laboratory source. Combined results for $q_S$ and the saturated value of the polarization are analyzed in terms of Landau theory previously developed to describe the para-/antiferro-/ferroelectric sequence of phase transitions in solid ferroelectrics. The analysis of $q_S$ indicates that the polarization modulation can be reasonably well approximated by a simple sinusoid in DIO, whereas in FNLC919 a sharp decrease in $q_S$ close to the antiferro- to ferroelectric transition indicates development of a soliton-like profile.
\end{abstract}

\maketitle
\newpage

\section{Introduction}

Since the first reports on the synthesis and characterization of highly polar liquid crystals exhibiting ferroelectricity in a fluid nematic state \cite{Mandle2017,Nishikawa2017,Mertelj2018,Chen2020,Sebastian2020}, there has been an explosion in discovery of related phases that constitute an ever expanding ``ferroelectric nematic realm"  \cite{Saha2021,Kikuchi2022,Chen20222,Chen2023,Kikuchi2024,Gibb2024,Song2024,Hobbs2024,Karcz2024,Adaka2024,Hobbs2025,Strachan2025,Pociecha2025,Hobbs20252,Guragain2025}. Among the most intriguing is an intermediate, antiferroelectric ($AF$) phase observed in certain compounds over a range of temperature between the ordinary, paraelectric nematic ($N$) and ferroelectric nematic ($N_F$) phases.

Key structural features of the intermediate phase were established in the ferroelectric nematic compound 2,3’,4’,5’-tetrafluoro[1,1’-biphenyl]-4-yl-2,6-difluoro-4- (5-propyl-1,3-dioxan-2-yl) benzoate (abbreviated DIO) \cite{Nishikawa2017} by Chen et al \cite{Chen2023}, who performed resonant carbon-edge and off-resonant small angle X-ray scattering (SAXS) experiments utilizing synchrotron sources. They demonstrated that the polarization field $\vec{P}$, which points along the axis of the nematic director $\hat{n}$, alternates with wavevector $\vec{q}_A$ in a direction normal to $\hat{n}$. The polarization modulation is accompanied by a weak, smectic-like mass density wave, with wavevector $\vec{q}_S = 2 \vec{q}_A$. These experimental results were preceded by theoretical \cite{Rosseto2020} work on a modulated splay nematic ($N_S$) phase that was initially assigned to the $N_F$ phase in the compound RM734 \cite{Mertelj2018} and by a separate study using Landau mean-field theory to identify the intermediate phase in DIO as a modulated, antiferroelectric phase \cite{Kats2021}.

The measured value of $q_S = 2 q_A \simeq 0.7$~nm$^{-1}$ in DIO implies a spatial period $\Lambda_A = 2 \Lambda_S \simeq 18$~nm for the polarization modulation, which varies only weakly over the $\sim 15^\circ$C range of the antiferroelectric phase. Synchrotron X-ray measurements of $\Lambda_S$ and $\Lambda_A$ on the nanometer scale, also with $\Lambda_A = 2 \Lambda_S$, were reported by other investigators \cite{Cruickshank2023} in a homologous series based on RM734 \cite{Mandle2017}, the compound in which the $N_F$ phase was first unambiguously identified \cite{Chen2020}.

The combination of nanoscale density and polarization modulations with commensurate wavevectors along a common axis perpendicular to $\hat{n}$ motivated the designation of a ``smectic-$Z_A$" ($SmZ_A$) phase, where ``$Z$" denotes tilt of $\hat{n}$ at $90^\circ$ angle relative to the layer normal $\hat{q}_S$ and ``$A$" refers to the antiferroelectric order. Observations of chevron deformation in the layer structure \cite{Chen2023}, together with studies of Grandjean textures in samples containing a chiral additive \cite{Thapa2024}, Freedericksz transition experiments \cite{Gleeson2025}, and light scattering measurements on layer/director fluctuations \cite{Ghimire2025}, indicate that the density/polarization modulation in DIO is one dimensional.

\begin{figure*}
	\includegraphics[width=1\textwidth]{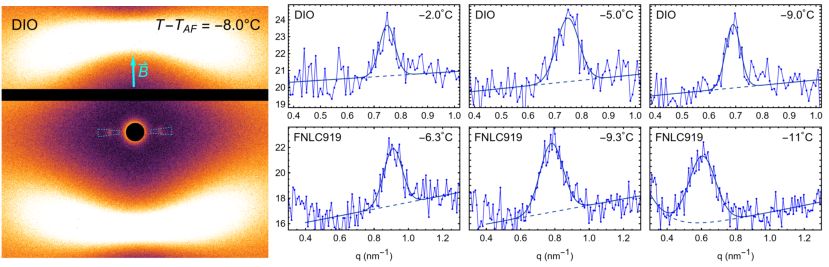}
	\caption{{\it At left:} Representative small-angle 2D X-ray diffraction pattern taken on DIO in the middle of the antiferroelectric phase (at temperature $T-T_{AF} = -8^\circ$C relative to the para- to antiferroelectric transition temperature). The direction of the magnetic field $\vec{B}$ aligning the collinear director and polarization fields is indicated by the cyan arrow. The pair of weak Bragg spots surrounded by dashed annuli correspond to the first harmonic of the smectic-$Z_A$ layering (period $8.9$~nm). {\it At right:} Representative plots of SAXS intensity vs scattering wavenumber $q$ in the antiferroelectric phase of DIO and FNLC919, after azimuthal averaging over the annular regions indicated on the SAXS pattern. The corresponding values of $T-T_{AF}$ are given in the upper right corners. The smooth lines are fits to a Gaussian lineshape, with peak value at wavenumber $q_S$, superposed on a slowly varying background function $B(q)$ shown by dashed lines (see Supplemental Material, sec.~S3).}
\end{figure*}
 
The present work reports on a quantitative comparison of the temperature dependence of the polarization magnitude and mass density wavenumber $q_S$ in two ferroelectric nematics -- the prototypical compound DIO and a proprietary mixture FNLC919 -- to a Landau theory originally developed to describe the para-/antiferro-/ferroelectric transitions in solid ferroelectrics \cite{Ishibashi1978,Qiu1986,Durand1986,Denoyer1986}. This theory was first invoked and applied to study ferroelectric nematics by Chen et al \cite{Chen2023} in their seminal paper on DIO. In addition to the variation of the polarization magnitude with temperature, the theory successfully describes two distinct limits in the behavior of $q_S = 2 q_A$ observed within the antiferroelectric phase: weakly temperature-dependent $q_S$ in DIO, corresponding to a nearly pure sinusoidal polarization modulation, and a sharply decreasing $q_S$ close to the antiferro- to ferroelectric transition in FNLC919, indicating an evolution from sinusoidal to soliton-like modulation. 

\section{Experimental details}

The liquid crystal DIO was synthesized at Kent State University and purified to isolate the {\it trans} isomer. FNLC919 was provided by Merck Electronics KGaA. Both materials exhibit the phase sequence $N$--$AF/SmZ_A$--$N_F$ on cooling with respective transition temperatures $T_{AF} = 84.0$ and $T_F = 67.7^\circ$C for DIO and $T_{AF} = 41.3$ and $T_F = 29.4^\circ$C for FNLC919. 

X-ray diffraction was conducted using a Xeuss 3.0 small/wide-angle x-ray scattering instrument (Xenocs, Inc) with a microfocus CuK$_\alpha$ source and a vacuum sample chamber equipped with a hot stage. The samples were contained in sealed, boron-rich glass capillaries ($2$~mm inside diameter, $10~\mu$m wall thickness). Two-dimensional diffraction patterns were recorded with a 2~hr acquisition time at fixed temperatures in the antiferroelectric phase after cooling from the paraelectric nematic phase, where the director was uniformly aligned by an applied magnetic field $\vec{B}$. The incident beam size was $0.9 \times 0.9$~mm$^2$.

Sandwich cells with syn-parallel rubbed, polyimide alignment layers were employed in the measurement of the polarization as a function of $T$. The sample thickness was $10~\mu$m, and the filled cells were epoxy sealed around the edges. An in-plane, alternating electric field (with triangular waveform and frequency $50$~Hz) was applied along the rubbing direction, and the induced polarization current associated with polarization reversal was recorded as a function of time. Typical fields required to saturate the current
varied from $0.1$~V/$\mu$m in the lower to $1$~V/$\mu$m in the upper antiferroelectric range. Further details are provided in the Supplemental Material (Figs.~S2, S3).

\section{Results}

Fig.~1 shows a small-angle 2D X-ray diffraction pattern taken on DIO at $T - T_{AF} = -8.0^\circ$C (near the middle of the antiferroelectric phase). The intense, diffuse arcs at larger angle and centered along the direction of $\vec{B}$ are associated with skewed ``cybotactic" clusters of molecules that are characteristic of short-range positional correlations observed in the $N$ and $N_F$ phases. Along the direction normal to $\vec{B}$, and at smaller angle, are a pair of weak Bragg spots detected only in the antiferroelectric phase that confirm an electron density modulation running normal to the director and polarization fields and demonstrate that smectic-$Z_A$ layering can be detected using a conventional X-ray source \cite{Paul2025}. The corresponding value of $q_S=0.706$~nm$^{-1}$ (period $\Lambda_S = 8.90$~nm) closely matches the value $0.705$~nm$^{-1}$ from the prior synchrotron X-ray measurements on DIO at the same temperature \cite{Chen2023}.
\begin{figure*}
	\includegraphics[width=.9\textwidth]{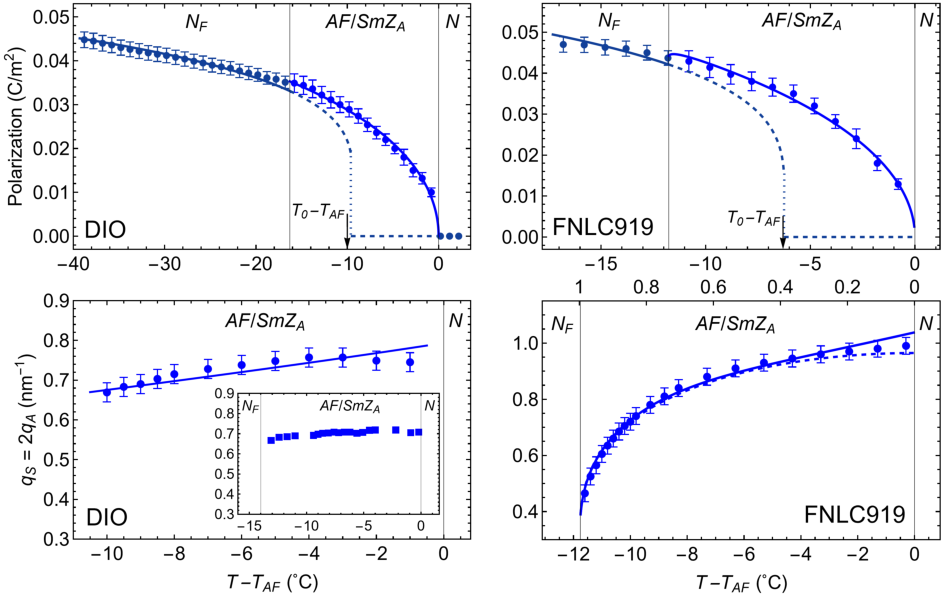}
	\caption{Temperature dependence of the saturated polarization and wavenumber $q_S = 2 q_A$ associated with the $SmZ_A$ density modulation in {\it trans} DIO (left column) and FNLC919 (right column). Data for the antiferroelectric phase are distinguished by the lighter blue color. For comparison, the inset in the bottom left panel plots $q_S$ data for DIO reported in ref.~\cite{Chen2023} and reprinted with permission. Solid curves through the data for DIO are plots of Eqs.~(3), (4), using a single set of Landau parameters, in the limit of a purely sinusoidal polarization profile over the antiferroelectric range (see Eqs.~(A4), (A5) in Appendix A). For FNLC919, the solid lines are plots of Eqs.~(3), (4), again for a single set of Landau parameters, in the case of polarization modulation that evolves from sinusoidal to soliton-like near the transition to the ferroelectric phase, where $q_S$ decreases sharply. The dashed line on the $q_S$ data is the quantity $1/K(k)$ in Eq.~(3) multiplied by a constant scale factor. In the top panels, dashed/dotted lines extrapolate the polarization in the ferroelectric phase modeled by Eq.~(4) through a ``virtual" first order ferro- to paraelectric nematic transition that would occur if the antiferroelectric phase was absent.}
\end{figure*}

Fig.~1 also displays representative plots of the scattered X-ray intensity ($I_s$) from the smectic-$Z_A$ density wave vs $q$ for selected temperatures. The intensity was averaged azimuthally over a $\pm 6^\circ$ range about the axis of the density wave, as indicated by dashed annuli in the SAXS pattern. Peaks at wavenumber $q_S$, though quite weak, are clearly distinguished above the background in Fig. 2. Fits of the peaks to a Gaussian profile, $A \exp[-(q-q_S)^2/\Delta q^2] + B(q)$, were used to obtain values of $q_S$. Here $B(q)$ is a smooth function that models the background (see sec.~S3 of the Supplemental Material).

The temperature dependence of $q_S$ for DIO and FNLC919 is plotted vs $T-T_{AF}$ in the bottom row of Fig.~2. In the case of DIO, the data only cover the upper $60$\% of the antiferroelectric range because diffraction from the density wave for $T-T_{AF} \lesssim -10^\circ$C using our conventional X-ray source was too weak to identify the peak position. Over the observable range, the value of $q_S$ decreases slightly with decreasing $T$. The synchrotron X-ray diffraction measurements on DIO in ref.~\cite{Chen2023}, reprinted with permission in the inset in Fig.~2, show a marginally weaker variation of $q_S$ with $T$ over nearly the complete range of the antiferroelectric phase. The median values in the inset and main figure are in good agreement. 

In the case of FNLC919, $q_S$ could be determined down to about $0.4^\circ$C above the antiferro- to ferroelectric transition ($T = T_F$). Closer to $T_F$ the peak merges into the edge of the undiffracted main beam, and there is no clearly defined local intensity maximum. The value of $q_S$ decreases slowly through the upper antiferroelectric range and then drops sharply close to the transition, a behavior quite distinct from DIO. The value $q_S = 0.99$~nm$^{-1}$ just below the transition from the nematic phase is $\sim 30$\% higher than in DIO and corresponds to a layer spacing $\Lambda_S = 6.3$~nm.

In order to observe diffraction from the weak smectic-$Z_A$ density wave, relatively wide settings of the beam collimation slits had to be used, which significantly degraded the peak-width resolution. Thus, the inverse FWHM obtained from the analysis of the peak profiles in Fig.~2 does not provide an accurate estimate of the correlated domain size. That said, within the lower third of the antiferroelectric phase in the FNLC919 sample, where $q_S$ decreases more sharply, the peak broadens by about 50\%, indicating diminishing smectic-$Z_A$ layer correlations as $T \rightarrow T_F$.

The top row of Fig.~2 shows experimental results for the saturated polarization in the antiferroelectric and ferroelectric phases of DIO and FNLC919. With decreasing temperature, the polarization magnitude increases continuously through the antiferroelectric phase, crossing into the ferroelectric phase without a noticeable jump.

\section{Discussion}

Various theoretical approaches have been put forward to explain the occurrence of an antiferroelectric phase between the para- and ferroelectric nematic phases. One approach \cite{Rosseto2020,Kats2021,Emelyanenko2022,Emelyanenko2024,Zou2024,Rupnik2025,Paik2025} focuses explicitly on the coupling of polar interactions with elastic deformations of the director field. It describes the antiferroelectric ordering by filling space with splayed polar domains that reverse polarization direction periodically along one or two directions (normal to the average director). The splay modulation is stabilized by a balance of Frank elastic energy, flexoelectric coupling between director splay and polarization $\vec{P}$, electrostatic energy incurred by accumulation of bound charge in the splayed domains, Landau terms in $P^2$ that favor a spontaneous polarization below a characteristic temperature, and gradient terms in $\vec{P}$. A ``minimal" model of the free energy density, given by Kats \cite{Kats2021}, anticipated the possibility of a stable modulated (antiferroelectric) phase between uniform para- and ferroelectric nematic phases. In the simplest splay-based models, the modulation period $\Lambda_A$ is set by competition between the intrinsic splay curvature (determined by the flexoelectric coupling constant, splay elastic constant, and polarization amplitude) and the electrostatic energy incurred by the accumulation of bound charge due to the polarization splay.

The electrostatic energy depends strongly on the number density $n_q$ and charge $q$ of ions (present as impurities or added as dopants) that effectively screen the bare Coulomb potential beyond the Debye length, $\lambda_D = \sqrt{\epsilon \epsilon_0 k_B T/(2 q^2 n_q)}$. After doping with a miscible ionic fluid (thereby reducing $\lambda_D$), certain ferroelectric nematics exhibit optical stripe patterns, consistent with a splay modulation, over a temperature range between the para- and ferroelectric nematic phase \cite{Rupnik2025}. The modulation period is in the micron range, in agreement with calculations based on a model in which domains of opposite polarization splay (but constant $| \vec{P} |$) are separated by well-defined domain walls across which the direction of $\vec{P}$ reverses. Both $\Lambda_A$ and the calculated width of the walls, $d_w$, significantly exceed the Debye length over the range of the antiferroelectric phase and over at least an order of magnitude change in $\lambda_D$ -- i.e., $\lambda_D \ll d_w < \Lambda_A$.

In contrast to the ion-doped ferronematic materials, the period of polarization modulation in the undoped DIO and FNLC919 samples studied in the present work is on the nanometer scale, $\Lambda_A = 2 \pi/q_A = 4 \pi/q_S \sim 10-20$~nm. In this case, if $\lambda_D \gg \Lambda_A$, the electrostatic energy cost due to the bare Coulomb potential strongly suppresses polarization splay over lengths $\sim \Lambda_A$. To estimate $\lambda_D$ for the DIO and FNLC919 samples studied, their ionic content was measured in the nonpolar $N$ phase; details are given in the Supplemental Material. The results are $n_q \approx 1.0 \times 10^{21}$~$e$/m$^3$ for DIO and $1.3 \times 10^{21}$~$e$/m$^3$ for FNLC919 at $T - T_{AF} = 40^\circ$C and $20^\circ$C, respectively. Taking $q = e$ and $\epsilon = \epsilon_\perp = 50$, and accounting for uncertainties in the values of $n_q$, give $\lambda_D \approx 220 \, (\pm 5\%)$~nm for the DIO and $\approx 170 \, (\pm 7.7\%)$~nm for the FNLC919 samples. These values are $\sim 10$ times larger than the respective values of $\Lambda_A$ determined from the data for $q_S$ and imply $\lambda_D \gg d_w$, opposite of the result obtained from the splay-modulated model applied to ion-doped ferroelectric nematics for a large range of $n_q$. 

Chen et al \cite{Chen2023} and collaborators \cite{Giesselmann2024} considered an alternative model of the antiferroelectric order for the case where ionic screening of polarization charge is weak, and polarization splay is expelled due to high electrostatic energy cost. The director $\hat{n}$ is essentially uniform, while the amplitude $P$ of the polarization $\vec{P} = P \hat{n}$ passes through zero and reverses sign periodically across Ising walls. In this so-called ``pure polarization reversal" (PPR) model, a weak mass density modulation is produced by slightly closer packing between molecules in the polarized domains relative to the domain walls where the polarization passes through zero. The smectic-like density wave is therefore a consequence of the antiferroelectric polarization modulation. (The same mechanism also applies in the splay-modulated models, where the domain walls are also Ising-type.)

To develop an analytic framework for the PPR model incorporating the para-/antiferro-/ferroelectric sequence of phase transitions, Chen et al insightfully drew a connection to the well-established Landau theory that describes analogous phenomena in solid ferroelectrics \cite{Ishibashi1978,Qiu1986,Durand1986,Denoyer1986}. They successfully used the theory to explain both the stabilization of the antiferroelectric/smectic-$Z_A$ phase by long-range dipolar interactions and the weak temperature dependence of the wavenumber $q_S = 2 q_A$ in DIO that follows when the modulation of $P$ is nearly sinusoidal. 

The remainder of this section demonstrates that the Landau theory invoked by Chen et al can be used to analyze quantitatively the results in Fig.~2 for the key equilibrium parameters -- polarization magnitude and wavenumber -- that characterize the antiferroelectric phase in both DIO and FNLC919, where the experimental temperature dependencies of $q_S$ are quite distinct. The equilibrium free energy density is expanded in even powers of $P$ and its derivatives along a direction $\hat{x}$ normal to the (fixed) polarization axis ($\hat{z}$) as
\begin{eqnarray}
	f = && \frac{a(T)}{2} P^2 + \frac{b}{4} P^4 + \frac{c}{6} P^6 + \frac{\alpha}{2} \left({d P} \over {d x}\right)^2 \nonumber \\ && + \frac{\beta}{2} \left({d^2 P} \over {d x^2}\right)^2 + \frac{\eta}{2} P^2 \left({d P} \over {d x}\right)^2,
\end{eqnarray}
where $a (T) = a_0 (T - T_0)$ and the coefficients $a_0, c,\beta,\eta$ are positive and constant in $T$. If $\alpha<0$ in Eq.~(1), an antiferroelectric phase with $P = P(x)$ alternating along $x$ may be stabilized between higher temperature para- and lower temperature ferroelectric phases. For $\alpha>0$, there is no stable antiferroelectric phase, and a direct, first (second) order para- to ferroelectric transition occurs at $T=T_0 + 3 b^2 / 16 a_0 c$ ($T = T_0$) if $b<0$ ($b>0$). The term in $\beta$ in Eq.~(1) stabilizes the antiferroelectric ordering at finite $q_A$. The term in $\eta$ produces an explicit dependence of $q_A$ on $P$ (see Eqs.~(A2) and (A4) in the Appendix).

Variation of the integrated $f$ in Eq.~(1) yields a nonlinear fourth order differential equation for $P$. As shown by Klepikov and Berezovsky \cite{Klepikov1996} (see Appendix A), approximate solutions are Jacobi elliptic functions, among which is the elliptic sine,
\begin{equation}
	P = P_0 \, \mathrm{sn} (lx, k) \equiv P_0 \, \mathrm{sn} (u, k) .
\end{equation}
Here $P_0$, $l$, and the modulus $k$ ($0 \leq k < 1$) are variational parameters which minimize $f$ averaged over the period $\Lambda_A$ of the $P$ profile, $\Lambda_A = 4 K(k)/l$ where $K(k)$ is the complete elliptic integral of the first kind. 

Eq.~(2) reduces to a simple sinusoid, $P = P_0 \sin (lx)$, when $k \rightarrow 0$, which is reasonable to expect close to a weakly first order antiferro- to paraelectric transition. As $k \rightarrow 1$ the periodic profile becomes increasingly soliton-like, with the characteristic width of Ising walls between domains wherein $P \approx \pm P_0$ becoming a smaller fraction of the spatial period. This behavior may be expected close to the ferroelectric phase. The period $\Lambda_A$ reduces to $2 \pi / l$ at $k = 0$ and diverges as $k \rightarrow 1$ (due to the divergence of $K(k)$), which implies a strong decrease in the wavenumbers $q_A$ and $q_S$, given by 
\begin{equation}
	q_S = 2 q_A = \frac{4 \pi}{\Lambda_A} = \frac{\pi \, l}{K(k)} 
\end{equation}
Such a decrease is observed in the results for $q_S$ at lower temperatures in the antiferroelectric phase of FNLC919 (Fig.~2) and, according to the model, signals the development of a soliton-like modulation. 

The $(k,T)$ dependence of $l$ and $P_0$ may be obtained by inserting Eq.~(2) into Eq.~(1), averaging over $x$, and minimizing the average free energy density $\bar{f}$ over $l$ and $P_0$. The result for $l$ is given by Eq.~(A2) in Appendix A. For $P_0$, one obtains
\begin{equation}
	P_0^2 (k,T) = -\frac{B(k)}{2 C(k)} + \sqrt{\frac{B(k)^2}{4 C(k)^2} - \frac{A(k,T)}{C(k)}} ,
\end{equation}
In the antiferroelectric phase, the quantities $A(k,T)$, $B(k)$, $C(k)$ are expressed (Appendix A, Eqs.~(A3)) in terms of the Landau coefficients $a(T)$, $b$, $c$, $\alpha$, $\beta$, $\eta$ and $u$-averaged combinations of $\mathrm{sn} (u, k)$ and its derivatives with respect to $u$. Stability of the antiferroelectric phase requires $C(k) > 0$. The nematic to antiferroelectric transition occurs for $\alpha < 0$ at the temperature ($T_{AF}$) where $A(k,T) = 0$. If $k \rightarrow 0$ as $T \rightarrow T_{AF}$ from below (pure sinusoidal polarization modulation at the para- to antiferroelectric transition), this condition gives $T_{AF} = T_0 + \frac{\alpha^2}{4 \beta a_0}$.

In the ferroelectric phase, $P$ is uniform in space and $A(k,T) \rightarrow a (T)$, $B(k) \rightarrow b$, $C(k) \rightarrow c$ in Eq.~(4). The transition from the antiferro- to ferroelectric phase happens at the temperature ($T_F$) where the value of $\bar{f}$ for the two phases is the same.
\begin{figure}
	[h!]
	\centering
	\includegraphics[width=.48\textwidth]{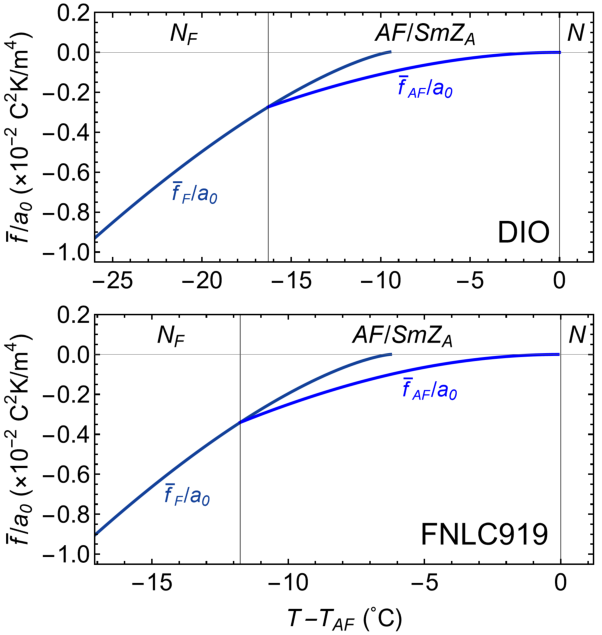}
	\caption{Average free energy densities (divided by $a_0$) calculated in the ferroelectric and antiferroelectric phases of DIO (top panel) and FNLC919 (bottom panel) from the model in Eqs.~(1) and (2), the expressions for $q_A$ and $P_0$ given in Eqs.~(3), (4), (A2)--(A5), and the values of reduced Landau coefficients and $T_0$ used to model the data in Fig.~2. The energies are equal at $T-T_{AF} = -16.3^\circ$C for DIO and $T-T_{AF} = -11.8^\circ$C for FNLC919 (indicated by vertical lines).}
	\end{figure}

The weak $T$ dependence of $q_S = 2 q_A$ in DIO implies that $K(k)$ in Eq.~(3) does not vary significantly over the antiferroelectric range. One may then consider simplifying the model by taking the small $k$ limit, where Eq.~(2) reduces to $P = P_0 \sin (q_A x)$ and $K(k \rightarrow 0) \rightarrow \pi/2$.
	
The solid lines in Fig.~2 through the data points for saturated polarization and $q_S$ in DIO are calculated from the $k \rightarrow 0$ limits of Eqs.~(3), (4), 
using a {\it single} set of reduced Landau coefficients for both the antiferro- and ferroelectric phases. Their values are $\frac{b}{a_0} = -4.36\times10^3$~m$^4$K/C$^2$, $\frac{c}{a_0} = 9.21\times10^6$~m$^8$K/C$^4$, $\frac{\alpha}{a_0} = -1.27\times10^{-16}$~m$^2$K, $\frac{\beta}{a_0} = 4.04\times10^{-34}$~m$^4$K, and $\frac{\eta}{a_0} = 1.69\times10^{-13}$~m$^6$K/C$^2$ with $T_0 - T_{AF} = -10^\circ$C. 

These values are constrained by the following conditions: (i) $T_0-T_{AF} = -\frac{\alpha^2}{4 \beta a_0}$; (ii) $\bar{f}_{AF} (T_F) = \bar{f}_F (T_F)$ (the average free energy densities of the ferro- and antiferroelectric phases are equal at the transition $T=T_F$); (iii) the amplitude of the modulated antiferroelectric polarization should be comparable to (or less than) the ferroelectric polarization at $T_F$; and (iv) $T_F < T_0 < T_{AF}$. As a result of these conditions, only four parameters in the above list may be freely varied. In the polarization measurement, $T_{AF}$ and $T_F$ were determined from the onset of antiferroelectric switching and the crossover from antiferroelectric to ferroelectric switching, respectively. In the SAXS data, the initial detection of the density modulation on cooling was used to fix $T_{AF}$ (within $\sim 0.5^\circ$C). Fig.~3 confirms that the condition $\bar{f}_{AF} (T_F) = \bar{f}_F (T_F)$ is satisfied for the parameters used to model the data in Fig.~2 for DIO. The vertical line in the figure corresponds to $T_F - T_{AF} = -16.3^\circ$C, which matches the experimentally determined value from the polarization data.
 
In contrast to the analysis for DIO, the $k$ dependence in Eq.~(3) becomes important to describe the behavior of $q_S$ in FNLC919 -- specifically, its sharp pretransitional downturn as $T \rightarrow T_F$. The dashed curve through the $q_S$ data points for FNLC919 (Fig.~2) is a plot of $1/K(k)$ on a linear scale in $k$ (top axis in the figure) multiplied by a constant factor to match the data, which are plotted on a linear scale in $T$ (bottom axis). The comparison suggests the linear function $k(T) = k_0 (T_{AF} - T)$ is a reasonable first order approximation for the $T$ dependence of $k$ over the measured range of $q_S$.

The solid lines plotted on the data for FNLC919 in Fig.~2 are calculated from Eqs.~(3), (4) using the set of reduced Landau coefficients $\frac{b}{a_0} = -6.78\times10^2$~m$^4$K/C$^2$, $\frac{c}{a_0} = 2.12\times10^6$~m$^8$K/C$^4$, $\frac{\alpha}{a_0} = -4.68\times10^{-17}$~m$^2$K, $\frac{\beta}{a_0} = 8.68\times10^{-35}$~m$^4$K, $\frac{\eta}{a_0} = 3.80\times 10^{-14}$~m$^6$K/C$^2$, and the linear form for $k(T)$ with $k_0 = 0.0848$. The value of $T_0 - T_{AF}$ is $-6.3^\circ$C. Again, only four of Landau parameters, plus $k_0$, are varied freely. At the lower end of the data plotted for $q_S$, the modulus $k = 0.98$, which from Eq.~(2) implies a soliton-like polarization profile at low temperature in the antiferroelectric phase of FNLC919. Despite deviations from the data for $q_S$ at higher $T$ and in the slope of $P$ in the $N_F$ phase, the agreement between the model and experimental results is fairly good. 

For the parameter values listed above, the plots of $\bar{f}_{AF}$ and $\bar{f}_F$ in Fig.~3 give $T_F - T_{AF} = -11.8^\circ$C in FNLC919. This value, obtained from the modeling and shown by the vertical lines in the right panels of Fig. 2, is still $\sim 0.2^\circ$C above the temperature at which the smectic-$Z_A$ diffraction peak fully disappeared into the shoulder of the undiffracted beam. The discrepancy may reflect inaccuracy in the assumption of a linear variation of $k$ with $T$ or possibly a narrow coexistence region due to the multi-component composition of FNLC919. Higher resolution SAXS data close to $T_F$ and a proper calculation of the temperature dependence of $k$ -- i.e., by minimizing $\bar{f}$ over $k$ and then solving the resulting transcendental equation to find the value of $k$ at each $T$ for a given set of Landau coefficients -- would enable a better test of the model for $q_S$ close to $T_F$, as well as possibly improving its description of the data near $T_{AF}$. One could also consider retaining powers of $P$ in Eq.~(A1) beyond $P^4$, which would result in more complicated, hyperelliptic functions as solutions for the spatial profile of $P$ \cite{Berezovsky1998}.

It is interesting that the model for the amplitude of the polarization modulation (Eq.~(4)) accurately represents the saturated polarization under applied electric field in the antiferroelectric phase of both materials studied. This {\it suggests} that under an applied field sufficient to saturate the polarization current, the polarization switches between oppositely polarized states with $|\vec{P}| \simeq P_0$ (the amplitude of the modulation in the ground state).

As a final point, detection of higher harmonics of the smectic-$Z_A$ layer structure by SAXS as $T \rightarrow T_F$ in FNLC919 would provide a more direct experimental indication of a developing soliton-like polarization modulation. However, this would be challenging for three reasons: very weak intrinsic scattering from the first harmonic, lower relative amplitudes of the higher harmonics (even approaching a strongly soliton-like profile), and suppression of the higher harmonic intensities by thermal fluctuations of the smectic-$Z_A$ layers. As described in sec.~S4 of the Supplemental Material, calculations using the model in Eq.~(2) confirm that the second harmonic is too weak to observe in the SAXS measurements reported here. 

\section{Summary}

To summarize, the magnitude and the wavenumber of the polarization field characterizing the antiferroelectric/smectic-$Z_A$ phase of two ferroelectric nematic liquid crystals were compared quantitatively, and overall favorably, to a Landau theory originally developed to describe antiferroelectric ordering in solid ferroelectrics. The comparison to theory indicates that the spatial profile of the modulation evolves rather differently with temperature in the two liquid crystals studied. In the compound DIO, the modulation maintains an essentially sinusoidal (single harmonic) profile deep into the antiferroelectric phase, whereas in the commercial mixture FNLC919 a sharp decrease in the modulation wavenumber indicates development of a significantly soliton-like polarization profile on approach to the antiferro- to ferroelectric transition.

The present work also demonstrates that diffraction from the very weak mass density wave accompanying the polarization modulation in the antiferroelectric phase of ferroelectric nematics can be detected with conventional laboratory SAXS instrumentation.

\begin{acknowledgments}
The research reported in this paper was supported by the National Science Foundation under Grants No. DMR-2210083 (MB, AG, M, JG, AJ, SS) and No. DMR-2341830 (PK, ODL). The material FNLC919 was provided by Merck Electronics KGaA, Darmstadt, Germany. The authors acknowledge access to the X-ray scattering and Organic Synthesis facilities at the Advanced Materials and Liquid Crystal Institute at Kent State University, which was financially supported by the National Science Foundation (Grant No. DMR-2017845), the State of Ohio (The Ohio Department of Higher Education Action Fund), and Kent State University. The authors thank X. Chen and the coauthors of ref.~\cite{Chen2023} for permission to reprint previously published data for the wavenumber $q_S$ in the bottom left panel of Fig.~2.
\end{acknowledgments}
	
\bibliography{DIO_919_Landau}

\clearpage

\appendix
\counterwithin{figure}{section}

\section{Further details on approximate solution of the model presented in Eq.~(1)}

The physically relevant solutions of the differential equation obtained from $\delta (\int d \vec{r} f) = 0$ are periodic in $x$ and can be expressed as a Fourier series, $P(x) = \sum_n P_n \sin [(2n+1)q_A x]$, with the even harmonics excluded since $P$ alternates ($P \rightarrow -P$) in each half period in the antiferroelectric phase -- i.e., $P(x+\Lambda_A/2) = P(x+\pi/q_A) = -P(x)$. The amplitudes $P_n$ and wavenumber $q_A$ are variational parameters that depend on $T$ and are determined by minimizing the average free energy density $\bar{f}$ over one period.

Klepikov and Berezovsky \cite{Klepikov1996} demonstrated that the solutions also satisfy the following nonlinear equation for the first derivative of $P$
\begin{equation}
	\left( {d P} \over {d x}\right)^2 = \sum_{n=0}^{\infty} A_{2n}  P^{2n} = A_0 + A_2 P^2 + A_4 P^4 + \cdots . 
\end{equation}
The coefficients $A_{2n}$ are another set of variational parameters whose values minimize $\bar{f}$. 

Limiting the number of terms in the sum on the right hand side of Eq.~(A1) leads to approximations for the polarization profile that are more physically realistic than truncating the Fourier series (which would produce a profile with unphysical ripples due to limiting the number of spatial frequencies). The Jacobi sine function (Eq.~(2)) is a solution for terms up through $P^4$. If higher powers of $P$ are retained the solutions are more complicated, hyperelliptic functions. The variational parameters $P_0$, $l$, and the modulus $k$ in Eq.~(2) are related to $A_0$, $A_2$, $A_4$ by $A_0 = P_0^2 l^2$, $A_2 = -l^2 (k^2+1)$, and $A_4 = l^2 k^2 / P_0^2$.

Minimizing $\bar{f}$ over parameter $l$ in the approximate solution, Eq.~(2), gives
\begin{equation}
	l^2 = -\frac{\alpha}{2 \beta} {\overline{\mathrm{sn}^{\prime 2} (u,k)} \over \overline{\mathrm{sn}^{\prime\prime 2} (u,k)}} - \frac{\eta P_0^2}{2 \beta} \, {\overline{\mathrm{sn}^2 (u,k) \, \mathrm{sn}^{\prime 2} (u,k)} \over \overline{\mathrm{sn}^{\prime\prime 2} (u,k)}} ,
\end{equation}
where primes indicate derivatives with respect to the variable $u$ and the averages $\overline{( \cdots )}$ are computed as  $\overline{( \cdots )} = \frac{1}{4 K(k)} \int_0^{4 K(k)} ( \cdots ) du$. Their ratios depend only weakly on modulus $k$. Note that with $(\beta, \eta) > 0$, $l$ is real valued only if $\alpha < 0$. 
(a requirement for existence of an antiferroelectric phase).

The explicit forms of the averages in Eq.~(A2) are
\begin{eqnarray*}
	&&\overline{sn^2 (u,k)} = {{K(k)-E(k)} \over {k^2 K(k)}} \\
	&&\overline{sn^4 (u,k)} = {{(2 + k^2) K(k) - 2 (1 + k^2) E(k)} \over {3 k^4 K(k)}}\\
	&&\overline{sn^6 (u,k)} = {{(8 + 3k^2 + 4k^4) K(k) - (8 + 7k^2 + 8k^4) E(k)} \over {15 k^6 K(k)}}\\
	&&\overline{sn^{\prime 2} (u,k)} = {{(1 + k^2) E(k) - (1 - k^2) K(k)} \over {3 k^2 K(k)}}\\
	&&\overline{sn^{\prime \prime 2} (u,k)} = \\
	&&~~~~~~~~~~~{{(7 - 18 k^2 + 11 k^4) K(k) - (7 - 22 k^2 + 7 k^4) E(k)} \over {15 k^2 K(k)}}\\
	&&\overline{sn^2 (u,k)\,sn^{\prime 2} (u,k)} = \\
	&&~~~~~~~~~~~~~~~~~~~{{2 (1 - k^2 + k^4) E(k) - (2 - 3 k^2 + k^4) K(k)} \over {15 k^4 K(k)}} ,
\end{eqnarray*}
where $K(k)$ and $E(k)$ are complete elliptic integrals of the first and second kind, respectively.

The amplitude $P_0$ of the polarization modulation appearing in Eq.~(A2) is given by Eq.~(4) with
\begin{eqnarray}
	&&A(k,T) = a_0 (T-T_0) \, \overline{\mathrm{sn}^2(u,k)} - \frac{\alpha^2}{4 \beta} {\left(\overline{\mathrm{sn}^{\prime 2} (u,k)} \right)^2 \over \overline{\mathrm{sn}^{\prime\prime 2} (u,k)}} \nonumber \\
	&&B(k) = b \, \overline{\mathrm{sn}^4(u,k)} - \frac{\eta \alpha}{\beta} \, {\overline{\mathrm{sn}^2 (u,k) \, \mathrm{sn}^{\prime 2} (u,k)} \,\,\, \overline{\mathrm{sn}^{\prime 2} (u,k)} \over \overline{\mathrm{sn}^{\prime\prime 2} (u,k)}} \nonumber\\
	&&C(k) = c \, \overline{\mathrm{sn}^6(u,k)} - \frac{3 \eta^2}{4 \beta} \, {\left(\overline{\mathrm{sn}^2 (u,k) \, \mathrm{sn}^{\prime 2} (u,k)} \right)^2 \over \overline{\mathrm{sn}^{\prime\prime 2} (u,k)}} , 
\end{eqnarray}

One could combine all of the above with Eqs.~(1)--(4) and then minimize $\bar{f}$ over $k$ to obtain a complicated transcendental equation for the temperature dependence of the modulus $k$. A numerical analysis of the result could be used potentially to improve the modeling of the data for FNLC919 in Fig.~2.

In the $k=0$ limit (pure sinusoidal polarization profile), Eq.~(3) and (4) reduce to
\begin{eqnarray}
	&&q_S (T) = 2 q_A (T) = \sqrt{-\frac{2 \alpha}{\beta}-\frac{\eta P_{0,AF}^2 (T)}{2 \beta}}\\
	&&P_{0,AF}^2 (T) = - \frac{3b - \frac{\eta \alpha}{\beta}}{5c - \frac{3 \eta^2}{8 \beta}} + \nonumber \\
	&&~~~~~~~~~~~\sqrt{\left( \frac{3b - \frac{\eta \alpha}{\beta}}{5c - \frac{3 \eta^2}{8 \beta}} \right)^2 +\frac{8 a_0 (T_{AF}-T)}{5c - \frac{3 \eta^2}{8 \beta}}},
\end{eqnarray}
in the antiferroelectric phase with $T_{AF} = T_0 + \frac{\alpha^2}{4 \beta a_0}$.
These expressions were used in the analysis of the data for DIO in Fig.~2.

\end{document}